\documentclass[11pt]{article}

\begin{document}

\title{\LARGE The GEMS Approach to Stationary Motions in the Spherically Symmetric Spacetimes}

\author{Hong-Zhi Chen$^{1}$\thanks{Email: hzhchen@pku.edu.cn}, Yu Tian$^{2}$\thanks{Email:
ytian@itp.ac.cn}, Yi-Hong Gao$^{2}$\thanks{Email:
gaoyh@itp.ac.cn}, Xing-Chang Song$^{1}$\thanks{Email:
songxc@pku.edu.cn}\\
\\
$^{1}$ School of Physics, Peking University,\\
Beijing 100871, P. R. China\\
$^{2}$ Institute of Theoretical Physics, Chinese Academy of Sciences,\\
P.O. Box 2735, Beijing 100080, P. R. China}
\date{}

\maketitle

\begin{abstract}
We generalize the work of Deser and Levin on the unified
description of Hawking radiation and Unruh effect to general
stationary motions in spherically symmetric black holes. We have
also matched the chemical potential term of the thermal spectrum
of the two sides for uncharged black holes.
\end{abstract}

\newpage

\section{Introduction}

It is well known that Hawking radiation \cite{Hawking:1974sw} and
Unruh effect \cite{Unruh:1976db} are closely related phenomena.
Recently, a unified description of the two effects began to
emerge. Given that any $D$-dimensional geometry has a higher
dimensional global embedding Minkowski spacetime (GEMS), Deser and
Levin \cite{Deser:1997ri,Deser:1998bb,Deser:1998xb}, found that
the temperature detected by the detectors outside the black hole
is the same as the temperature detected by the embedded observers
in the corresponding GEMS. In their paper, they considered the
static observers in various spherically symmetric spacetimes, and
the corresponding observers in the GEMS are the usual Rindler
observers, i.e. the observers with constant linear
acceleration.\footnote {The GEMS approach has also been detailedly
studied by many other authors. See \cite{HKP,HKKP,HKOP}, for
example.} This correspondence is interesting and profound, and it
is worthwhile to extend the discussion to more general detectors,
not just the static ones. In this paper, we will generalize this
correspondence to general stationary motions in the static
spherically symmetric black holes and show that, not only the
temperature of the two sides are the same, but also are the whole
thermal spectra, including the chemical potential for uncharged
black holes.

The outline of the paper is as follows. First, we give the
derivation for the Schwarzschild black hole. Then we generalize it
to the more complicated RN-AdS black hole. And finally we discuss
some related issues.

\section{The Schwarzschild Black Hole}
We only consider the four dimensional cases. We first consider the
simplest Schwarzschild black hole, and then generalize it to more
general spherically symmetric case. Generally, a static
spherically symmetric black hole has four Killing vectors, one is
timelike corresponding to the time-translational invariance and
the other three are spacelike corresponding to rotational
invariance. In the previous works, they only considered detectors
following the integral curves of the timelike Killing vector field
$\partial_t$, i.e. static detectors. In this case, the same
detector in the GEMS will be in a motion of constant linear
acceleration, i.e. a standard Rindler detector. We want to
consider a more general case in which the motion of the detectors
are stationary in the sense that they follow the integral curves
of a general timelike Killing vector field, as thus the
corresponding motion in the GEMS is also a stationary one, which
we can deal with. In the case of the Schwarzschild black hole, it
is a circular motion with constant $r$ and $\theta$ in the usual
coordinates.

To verify the correspondence, we have to compute the thermal
spectra from the two sides and see if they are the same. We begin
with the black hole side.

\subsection{The Black Hole Side}
The Schwarzschild metric is
\begin{equation}\label{Sch}
ds^2=\Big(1-\frac{2M}{r}\Big)dt^2-\Big(1-\frac{2M}{r}\Big)^{-1}dr^2-r^2
(d\theta^2+\sin^2\theta d\phi^2).
\end{equation}
We consider the observer following the trajectory
\begin{equation} \label{detector}
r=r_0>2M,\quad \theta=\theta_0,\quad \phi=\Omega t,
\end{equation}
where $r_0$, $\theta_0$ and $\Omega$ are constants.\footnote{We
take $\Omega>0$, without loss of generality. And, of course, we
should take $\Omega r_0\sin\theta_0<1$ to avoid unphysical
motions.} The authors of \cite{Deser:1998xb} only considered the
case $\Omega=0$. To determine the spectrum observed by such an
observer, we can not use the usual methods directly because the
observer is moving while the usual methods are applied to a static
observer. However, we can make the following coordinate
transformation to the observer's rest frame,
\begin{equation}
\tilde{\phi}=\phi-\Omega t.
\end{equation}
In this new frame, the metric becomes (taking
($x^0,x^1,x^2,x^3)=(t,r,\theta,\tilde{\phi})$)
\begin{eqnarray}\label{rotatesch}
ds^{2} &=&
\Big(1-\frac{2M}{r}-r^2\Omega^2\sin^2\theta\Big)dt^2-\Big(1-\frac{2M}{r}\Big)^{-1}dr^2
\nonumber \\
&& -\;r^2 (d\theta^2+\sin^2\theta
d\tilde{\phi}^2)-2r^2\Omega\sin^2\theta d\tilde{\phi}dt.
\end{eqnarray}
Note that this metric is stationary and axisymmetric. The detector
with trajectory (\ref{detector}) is now at the position of
\begin{equation}\label{rest}
r=r_0>2M,\quad \theta=\theta _0,\quad \tilde{\phi}=0.
\end{equation}
There are different methods to calculate the Hawking temperature
of a rest detector in this case. We will apply the method of
Damour and Ruffini \cite{Damour:1976jd}.\footnote{See also
\cite{zhao} for extensive discussions of D-R method.} First, from
the condition
\begin{equation}
\hat {g}_{00}\equiv g_{00}-\frac{g^2_{03}}{g_{33}}=0,
\end{equation}
it is easy to see that the horizon of metric (\ref{rotatesch}) is
at
\begin{equation}
  r_\mathrm{h}=2M.
\end{equation}
And from the following formula
\begin{equation} \label{surgravity}
\kappa=\lim_{\hat{g}_{00} \rightarrow
0}(-\frac{1}{2}\sqrt{\frac{g^{11}}{-\hat{g}_{00}}}\hat{g}_{00,1}),
\end{equation}
we get the surface gravity of the horizon:
\begin{equation}
\kappa=\frac{1}{4M}. \end{equation} Both of these two quantities
are the same as those of the usual Schwarzschild black hole, as
expected.

We then consider the massless Klein-Gordon equation in this
spacetime, which is
\begin{equation}\label{klein}
\frac{1}{\sqrt{-g}}\frac{\partial}{\partial x^{\mu}}
\Big(\sqrt{-g}g^{\mu \nu}\frac{\partial\phi}{\partial
x^\nu}\Big)=0.
\end{equation}
We separate the variables as follows,
\begin{equation}\label{sep}
\phi(t,r,\theta,\tilde{\phi})=e^{-{\rm i}(\omega
t-m\tilde{\phi})}\chi (\theta) \psi(r),
\end{equation}
where $\omega$ and $m$ is the coordinate energy and the magnetic
quantum number of the particles, respectively. We further define
the Regge-Wheeler radial coordinate,
\begin{equation}\label{tortoise}
r_{*}=r+2M\ln\frac{r-2M}{2M},
\end{equation}
and put eqs.(\ref{sep},\ref{tortoise}) into eq.(\ref{klein}).
After separating the variables, we finally obtain the radial
equation
\begin{equation}
\frac{d^2\psi(r_*)}{dr^2_*}+\frac{4(r-2M)}{r^2}\frac{d\psi(r_*)}{dr_*}
+[(\omega+m\Omega)^2+\frac{2M-r}{r^3}\lambda]\psi(r_*)=0,
\end{equation}
where $\lambda$ is the constant from separating variables. When
taking the limit $r\rightarrow r_\mathrm{h}$, this equation
becomes
\begin{equation}
\frac{d^2\psi(r_*)}{d r^2_*}+(\omega+m\Omega)^2\psi(r_*)=0.
\end{equation}
Following \cite{Damour:1976jd}, we deduce that the detector will
detect a thermal spectrum of the form
\begin{equation}
N_\omega=\frac{1}{e^{(\omega+m\Omega)/T_0}-1},
\end{equation}
where
\begin{equation}
T_0=\frac{\kappa}{2\pi}=\frac{1}{8\pi M },
\end{equation}
and $N_\omega$ is the mean number of particles with energy
$\omega$. However, we note that the temperature $T_0$, energy
$\omega$ and angular velocity $\Omega$ are all coordinate
quantities. To get the corresponding local quantities detected by
the detector at position (\ref{rest}), we just divide them by the
red shift factor from eq.(\ref{rotatesch}), and finally get
\begin{equation}\label{bhspec}
N_{\tilde{\omega}}=\frac{1}{e^{(\tilde{\omega}+m\tilde{\Omega})/T}-1},
\end{equation}
where now
\begin{equation}\label{omega}
T=\frac{T_0}{f}=\frac{1}{8\pi M f},\quad
\tilde{\omega}=\frac{\omega}{f},\quad
\tilde{\Omega}=\frac{\Omega}{f},
\end{equation}
and
\begin{equation}\label{redshit}
f(r_0,\theta_0,\Omega)=\sqrt{g_{00}}=\sqrt{1-\frac{2M}{r_0}-r^2_0\Omega^2\sin^2\theta_0}.
\end{equation}
Of course, the parameters $r_0$, $\theta_0$ and $\Omega$ are
arbitrary, and this formula applies to any detector with
trajectory (\ref{detector}).

\subsection{The GEMS Side}
The above Schwarzschild space can be embedded in a flat $D=6$
Minkowski spacetime with metric
\begin{equation}\label{6dm}
ds^2=(dz^0)^2-(dz^1)^2-(dz^2)^2-(dz^3)^2-(dz^4)^2-(dz^5)^2,
\end{equation}
as follows \cite{fronsdal}:
\begin{eqnarray}
z^0 &=& 4M\sqrt{1-\frac{2M}{r}}\sinh\Big(\frac{t}{4M}\Big), \nonumber \\
z^1 &=& 4M\sqrt{1-\frac{2M}{r}}\cosh\Big(\frac{t}{4M}\Big), \nonumber \\
z^2 &=& \int dr \sqrt{\frac{2Mr^2+4M^2r+8M^3}{r^3}}, \nonumber \\
z^3 &=& r\sin\theta \sin\phi, \nonumber \\
z^4 &=& r\sin\theta \cos\phi, \nonumber \\
z^5 &=& r\cos\theta. \label{schembed}
\end{eqnarray}
It should be emphasized that this embedding can be extended to
cover $r<2M$.\footnote{The issue of the complete global embedding
had been discussed in \cite{Hong:2003xz}.} We see that a detector
following the trajectory (\ref{detector}) will effectively be an
ordinary 4-dimensional Rindler motion superposed with a circular
motion whose plane is perpendicular to the acceleration of the
Rindler motion. Because when $\Omega=0$, the results of
\cite{Deser:1998xb} indicate that the Unruh detector detects the
same temperature as that of the Hawking detector in the
Schwarzschild space, intuitively, we expect this detector will
detect the same spectrum as eq.(\ref{bhspec}) when $\Omega\neq0$,
since both sides we are considering now are superposed with the
same circular motion. Although this is not a standard Rindler
motion, we can still apply the method of Bogoliubov transformation
in \cite{Unruh:1976db} to calculate the spectrum detected by such
a detector. The quantization of field theory in general stationary
coordinate systems had been discussed previously in
\cite{Letaw:1980yv,Letaw:1980ik}. Their discussions have been
extended in \cite{Korsbakken:2004bv}, including detailed
discussions on the causal structures. To apply the results of
\cite{Korsbakken:2004bv}, we have to determine what their
parameters $a$ and $\omega$ correspond to in our case.

First, we briefly summarize their results relevant to us. It is
easy to see that our case corresponds to $a\neq0$, $\omega_x\neq0$
and $\omega_z=0$ in their paper, where $a$ is the proper
acceleration of the detector along $x$-direction and $\omega_x$ is
the proper angular velocity of the detector in the plane
perpendicular to $x-$ axis. For such a motion, the trajectory is
easily found to be
\begin{eqnarray}
t' &=& C\sinh (a\tau), \nonumber \\
x' &=& C\cosh (a\tau)+D, \nonumber \\
r' &=& r_0, \nonumber \\
\phi' &=& \omega_x\tau, \label{noembed}
\end{eqnarray}
where $t'$, $x'$, $r'$ and $\phi'$ are the cylindrical coordinates
in the 4-dimensional Minkowski space, $C$, $D$ and $r_0$ are
constants, and $\tau$ is the proper time of the detector. In this
case, there are horizons in the detector's rest frame and the
vacuum state is defined by the standard canonical quantization
procedure in this frame. Unlike in the standard Rindler motion,
the restriction of the positive norm modes to one side of the
horizon does not at the same time restrict the positive frequency
modes to the same side. So the vacuum thus defined will not be the
true ground state and will have negative energy excitations. They
found that the detector will detect a thermal spectrum in the
Minkowski vacuum state and the spectrum is of the form
\begin{equation}\label{spectrum}
N_\varepsilon=\frac{1}{e^{(\varepsilon+m\omega_x)/T_\mathrm{e}}-1},
\end{equation}
where
\begin{equation}
T_\mathrm{e}=\frac{a}{2\pi},
\end{equation}
$m$ is the magnetic quantum number of the particle, and
$\varepsilon$ is the local energy.

Now, we come to our embedding space. For a detector following the
trajectory (\ref{detector}), the metric (\ref{6dm}) becomes
\begin{equation}
ds^2=\Big(1-\frac{2M}{r_0}-r^2_0\Omega^2\sin^2\theta_0\Big)dt^2,
\end{equation}
so we see that $t$ is not the proper time. However, it is easily
converted to the proper time by a rescaling
\begin{equation}
t=\frac{\tilde{t}}{f},
\end{equation}
where $f$ is defined in eq.(\ref{redshit}). After such a
rescaling, the trajectory of the embedded detector
becomes\footnote{Here we have combined $z^3$ and $z^4$ into polar
coordinates $R$ and $\Phi$.}
\begin{eqnarray}
z^0 &=& 4M\sqrt{1-\frac{2M}{r_0}}\sinh\Big(\frac{\tilde{t}}{4M f}\Big), \nonumber \\
z^1 &=& 4M\sqrt{1-\frac{2M}{r_0}}\cosh\Big(\frac{\tilde{t}}{4M f}\Big), \nonumber \\
R &=& {\rm constant}, \nonumber \\
\Phi &=& \Omega'\tilde{t}, \nonumber \\
z^2 &=& {\rm constant},\quad z^5={\rm constant}, \label{embed}
\end{eqnarray}
where
\begin{equation}\label{omega'}
\Omega'=\frac{\Omega}{f}.
\end{equation}
Comparing eq.(\ref{omega}) and eq.({\ref{omega'}), we see that
\begin{equation}
\tilde{\Omega}=\Omega'.
\end{equation}
From eq.(\ref{noembed}) and  eq.(\ref{embed}), we see that in our
case
\begin{equation}
a=\frac{1}{4M f},\quad \omega_x=\Omega'=\tilde{\Omega}.
\end{equation}
So from eq.(\ref{spectrum}), this detector will detect the thermal
spectrum
\begin{equation}
N_\varepsilon=\frac{1}{e^{(\varepsilon+m\Omega')/T_\mathrm{e}}-1}
\end{equation}
with
\begin{equation}
T_\mathrm{e}=\frac{a}{2\pi}=\frac{1}{8\pi M f}.
\end{equation}
Comparing} these results with those of Schwarzschild space
eqs.(\ref{bhspec},\ref{omega}), we find that they are identical.
Thus, we have verified that the correspondence is valid in this
more general case, including the chemical potential term.

\section{The RN-AdS Case}

We will now consider the more general spherically symmetric space,
namely RN-AdS black hole.\footnote{We restrict our discussion to
the non-extremal case.} The metric of 4-dimensional RN-AdS space
is \cite{carter}
\begin{equation}
ds^2=F(r,M,Q,R)dt^2-F^{-1}(r,M,Q,R)dr^2-r^2
(d\theta^2+\sin^2\theta d\phi^2),
\end{equation}
where
\begin{equation}
F(r,M,Q,R)=1-\frac{2M}{r}+\frac{Q^2}{r^2}+\frac{r^2}{R^2}.
\end{equation}
The outer horizon $r=r_\mathrm{h}$ of this space is determined
from
\begin{equation}
F(r_\mathrm{h},M,Q,R)=0,
\end{equation}
and the corresponding surface gravity $\kappa$ is
\begin{equation}
\kappa=\frac{r_\mathrm{h}^2-Q^2+3r_\mathrm{h}^4
R^{-2}}{2r_\mathrm{h}^3}
\end{equation}
The authors of \cite{Kim:2000ct} had discussed the GEMS of RN-AdS
space. They found that it can be embedded in a flat $D=7$ space
with metric
\begin{equation}
ds^2=(dz^0)^2-\sum\limits_{i=1}^{5}(dz^i)^2+(dz^6)^2,
\end{equation}
as follows:
\begin{eqnarray}\label{rnembed}
z^0 &=& \kappa^{-1}\sqrt{F(r,M,Q,R)}\sinh(\kappa t), \nonumber \\
z^1 &=& \kappa^{-1}\sqrt{F(r,M,Q,R)}\cosh(\kappa t), \nonumber \\
z^2 &=& z^2(r,r_\mathrm{h},Q,R), \nonumber \\
z^3 &=& r\sin\theta \cos\phi, \nonumber \\
z^4 &=& r\sin\theta \sin\phi, \nonumber \\
z^5 &=& r\cos\theta, \nonumber \\
z^6 &=& z^6(r,r_\mathrm{h},Q,R),
\end{eqnarray}
where $z^2$ and $z^6$ are complicated functions of the arguments
displayed whose explicit forms do not concern us.\footnote{Again,
see \cite{Hong:2003xz} for extending the embedding to
$r<r_\mathrm{h}$ case.} Note that this embedding is very similar
to that of the Schwarzschild space (\ref{schembed}), except for
the extra time dimension $z^6$ which is unimportant for the case
we will consider. Consider the following trajectory of a detector:
\begin{equation}\label{rn}
r=r_0>r_\mathrm{h},\quad \theta=\theta_0,\quad \phi=\Omega t,
\end{equation}
where $r_0$, $\theta_0$ and $\Omega$ are constants. We see from
eq.({\ref{rnembed}) that the embedded motion again reduces
effectively to a 4-dimensional one with $z^2$, $z^5$, $z^6$ being
constants. From the similarity of the metric and embedding of the
RN-AdS space with those of the Schwarzschild space, we can easily
show that the local temperature $T$ calculated form the two sides
are the same, and both are
\begin{equation}
T=\frac{\kappa}{2\pi f},
\end{equation}
where
\begin{equation}
f(r_0,\theta_0,\Omega)=\Big(1-\frac{2M}{r_0}+\frac{Q^2}{r^2_0}
+\frac{r^2_0}{R^2}-r^2_0\Omega^2\sin^2\theta_0\Big)^{\frac{1}{2}}.
\end{equation}
While for the chemical potential term, the issue is more subtle.
Because the RN black hole is charged, it will induce a chemical
potential term in the thermal spectrum even for a static detector.
However, on the GEMS side as discussed already by
\cite{Deser:1998xb}, it is a standard Rindler motion and the
detector will not detect a chemical potential in the thermal
spectrum. Thus even in that case, the chemical potential is not
properly matched. We found that this is always the case for
charged black hole because the embedding does not map all the
information of the black hole into the higher dimensional
spacetime, namely the electric field of the black hole does not
map to anything in the higher dimensional space. Obviously, to
match the chemical potential of the two sides in the case of
charged black hole properly, we have to settle this problem first.

\section {Concluding Remarks}
We have generalized the unified description of Hawking radiation
and Unruh effect proposed by Deser and Levin to the general
stationary motions. For the case of uncharged black hole, we also
matched the chemical potential of the two sides. Our discussion is
also easily extended to pure dS/AdS space. The issue of matching
the chemical potential for the charged black hole needs further
investigations to find a way of properly mapping all the
information contained in the charged case to the higher
dimensional embedding space. This unified picture of the two
effects seems to be two points of view of the same thing. The
detector is the same one, just  the points of view are different.
In one point of view it  moves in a curved spacetime and will
detect Hawking radiation associated with the horizon due to
gravity, while in the other one it moves in a higher dimensional
Minkowski space (possibly with more times) and will detect
radiation associated with the horizon due to acceleration. From
the previous calculations, we see that the key of the unification
is the correspondence of the quantum states of the two sides. For
example, in the Schwarzschild case, the detector is taken to be in
the Hartle-Hawking vacuum; while in the GEMS side, the detector is
taken to be in the Minkowski vacuum. It is this mapping of the two
states that makes the two points of view gives the same results.
Our generalization implies that this unified description is not
just a coincidence and there may be something important in it. The
correspondence between a lower dimensional quantum effect and a
higher dimensional quantum effect may also have something to do
with the holographic principle. The deep reason of this
correspondence deserves further study.

\section*{Acknowledgements}
The authors would like to thank Professors R.-G. Cai, B. Chen,
H.-Y. Guo, C.-G. Huang,  Z. Xu, Z. Zhao, Doctors Y. Ling and J.-B.
Wu for helpful discussions. This work is partly supported by NSFC
under Grants No. 10347148.

\end{document}